\documentclass[runningheads]{llncs}
\usepackage[T1]{fontenc}
\usepackage{graphicx}
\usepackage{epstopdf}
\usepackage{amsmath}
\usepackage{subfigure}
\usepackage{multirow}
\usepackage{amsmath,bm}
\usepackage{subfig}
\usepackage{caption}
\usepackage{amsfonts}
\usepackage{caption}
\usepackage{cite}

\begin{document}
\title{JCCS-PFGM: A Novel Circle-Supervision based Poisson Flow Generative Model for Multiphase CECT Progressive Low-Dose Reconstruction with Joint Condition}
\author{Rongjun Ge\inst{1,2}, Yuting He\inst{3}, Cong Xia\inst{4}, Yang Chen\inst{3}, Daoqiang Zhang\inst{1}, \\Ge Wang\inst{2}}

\institute{College of Computer Science and Technology, Nanjing University of Aeronautics and Astronautics, Nanjing, China\\
\and
Department of Biomedical Engineering, Rensselaer Polytechnic Institute, Troy, NY, USA\\
\and
Laboratory of Image Science and Technology, School of Computer Science and Engineering, Southeast University, Nanjing, China\\
\and
Jiangsu Key Laboratory of Molecular and Functional Imaging, Department of Radiology, Zhongda Hospital, Medical School of Southeast University, Nanjing, China
}
%
%}
\maketitle              % typeset the header of the contribution

\begin{abstract}
\vspace*{-1\baselineskip}
Multiphase contrast-enhanced computed tomography (CECT) scan is clinically significant to demonstrate the anatomy at different phases. In practice, such a multiphase CECT scan inherently takes longer time and deposits much more radiation dose into a patient body than a regular CT scan, and reduction of the radiation dose typically compromise the CECT image quality and its diagnostic value.
With \textbf{J}oint \textbf{C}ondition and \textbf{C}ircle-\textbf{S}upervision, here we propose a novel \textbf{P}oisson \textbf{F}low \textbf{G}enerative \textbf{M}odel (JCCS-PFGM) to promote the progressive low-dose reconstruction for multiphase CECT. JCCS-PFGM is characterized
by the following three aspects:
1) a progressive low-dose reconstruction scheme to leverages the imaging consistency and radiocontrast evolution along former-latter phases, so that enormously reduces the radiation dose needs and improve the reconstruction effect, even for the latter-phase scanning with extremely low dose;
% 前期与后期间的 和 对比增强 evolution， 从而逐期大大降低辐射剂量并提高重建效果for 极低剂量的后期。
2) a circle-supervision strategy embedded in PFGM to enhance the refactoring capabilities of normalized poisson field learned from the perturbed space to the specified CT image space, so that boosts the explicit reconstruction for noise reduction; % instead of CT-similar space image generation.
%2) a circle-supervision strategy embedded in PFGM to enhance the refactoring capabilities of normalized poisson field learned from the perturbed space to the specified CT image space, so that promote the explicit reconstruction instead of CT-similar space image generation.
% 强化从噪声空间到图片空间的映射.... 明确
%2) a circle-supervision based PFGM to 强化从噪声空间到图片空间的映射.... 明确,消除方向学习和。。。间的鸿沟
3) a joint condition mechanism to explore correlation between former phases and current phase, so that extracts the complementary information for current noisy CECT and guides the reverse process of diffusion jointly with multiphase condition for optimal recovery of clinically relevant structures.
%3) a joint condition to  有效地融合前期重建结果和当前期 以引导当前期更低剂量CT 的重建。
Our extensive experiments are performed on a clinical dataset consisting of 11436 images. The results show that our JCCS-PFGM achieves promising PSNR
up to 46.3dB, SSIM up to 98.5\%, and MAE down to 9.67 HU averagely on phases I, II and III, in quantitative evaluations, as well as gains high-quality readable visualizations in qualitative assessments.
All of these findings reveal our method a great potential to be adapted for clinical CECT scans at a much-reduced radiation dose.

%\keywords{Multiphase CECT\and Progressive low-dose reconstruction \and PFGM \and Circle-supervision \and Joint condition}
\end{abstract}
\section{Introduction}
\begin{figure}[t]
\centering
\includegraphics[width=1\textwidth,height=0.45\textwidth]{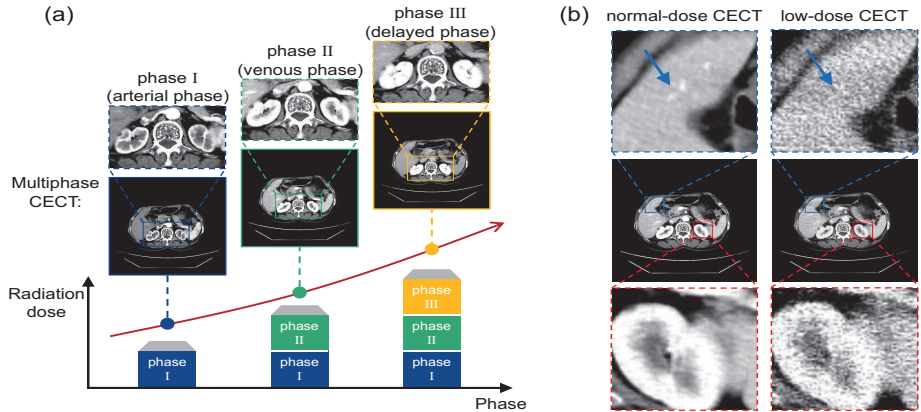}%,height=0.55\textwidth
\vspace*{-0.5\baselineskip}
\caption{Accumulated radiation dose and compromised image quality of low-dose CECT scans.
(a) Multiphase CECT scans comprehensively enable the demonstration of anatomy and lesion with the contrast agent evolution at different phases, but inherently lead to the accumulation of huge radiation dose for patients.
(b) Low-dose CECT images reconstructed using the standard algorithm are much more difficult to read than the normal-dose CECT images.
\vspace*{-1.5\baselineskip}
}\label{fig1}
\end{figure}

\vspace*{-0.5\baselineskip}
The substantial reduction of scanning radiation dose and its accurate reconstruction are of great clinical significance for multiphase contrast-enhanced computed tomography (CECT) imaging which demands much higher radiation dose than a regular CT examination.
1) Multiphase CECT requires multiple scans at different phases, which are typically called arterial phase, venous phase, delayed phase, and so on, to demonstrate the anatomy and lesion with the contrast agent evolution intra human body over time \cite{Meng2020}.  However, such multiphase CT scans inherently lead to the accumulation of huge radiation dose which should be minimized because of the potential risk of ionizing radiation to the patient \cite{Brenner,Rastogi}. As shown in Fig.~\ref{fig1}(a), during a triple-phase CECT scanning, the radiation dose absorbed by the patient is roughly three times that of the single phase. According to the ``as low as reasonably achievable'' (ALARA) principle \cite{Prasad2004}, it thus extremely urgent to greatly reduce the radiation dose and risk for clinical multiphase CECT examination.
2) However, the low-dose acquired CT image also exists the problems of noise interference, structural distortion and other problems. As enlarged region shown in Fig.~\ref{fig1}(b), the low-dose CECT images reconstructed using the standard algorithm have a much lower signal-to-noise ratio than normal-dose CECT counterpart. It brings great difficulty to read the anatomical structure with high noise, especially for inexperienced radiologist. Therefore, it is highly desirable to develop a dedicated algorithm for high-quality reconstruction in the scenario of multi-phase low-dose CECT.
%According to ``as low as reasonably achievable'' (ALARA) principle,  for radiation risk reduction,
%risk reduction.

%根据“”有效降低患者辐射是极其重要的，尤其对于多期扫描。
%1) 多期增强CT需要在不同期相如动脉期、静脉期、延迟期等进行多次扫描以获得。。。

As far as we know, most of the existing CT denoising methods focus on the single-phase low-dose CT (LDCT) reconstruction. Chen et al.\cite{Chen2017_1} trained a deep convolutional neural network (CNN) to transform LDCT images towards normal-dose CT images in a patch by patch fashion. In \cite{Chen2017}, a shortcut-aided symmetrical CNN was used to predict the noise distribution in LDCT. Shan et al.\cite{Shan2018} attempted to transfer a trained 2D CNN to a 3D counterpart for low-dose CT image denoising. In \cite{Ma2021}, an attention residual dense network was developed for LDCT
sinogram denoising. In \cite{Yin2019}, sinogram- and image- domain networks were trained for LDCT denoising in a progressive way. Zhang et al.\cite{CLEAR2021} further connected sinogram- and image- domain networks together for joint training. In \cite{Ye2022} and \cite{Ge2022}, parallel network architectures were put forward for dual-domain information sharing and mutual optimization.

Up to now, multi-phase LDCT reconstruction remains a largely unexplored area, although single-phase methods behave promising results on their issues \cite{Chen2017_1,Chen2017,Shan2018,Ma2021,Yin2019,CLEAR2021,Ye2022,Ge2022}. Due to multiple scans in a short time, it faces the inherent challenges: 1) Severe noise pollution is caused by the higher requirement of using much lower scanning dose to decrease multiphase radiation accumulation, compared to the single-phase imaging. Thus, it is extremely nontrivial to map from a lower-dose CECT scan with a significantly higher noise level to the normal-dose CECT counterpart. 2) Complex multiphase correlation with redundant information and interfering factors is induced by the evolution of contrast agent in the human body. Furthermore, behind the contrast perfusion and diffusion phenomena, strong causality also intrinsically exists among multiphase process. But how to deeply explore such consistency and evolution along the multiphase for further reducing the dose of later phase and improving imaging quality is still an open challenge. Clearly, an advanced solution is urgently needed for minimized radiation dose and maximized image quality.

%多期需要降低更多噪声，由于多期辐射剂量的积累。这就对低剂量重建提出更高的要求，以提取从更低剂量的CT with 更严重的噪声到NDCT的映射关系。Furthermore, 多期间存在强烈的相互关系，如何充分利用多期间一致性并挖掘evolution以进一步降低later phase 的剂量并提升成像质量仍然是一个open challenge.

In this paper, with \textbf{J}oint \textbf{C}ondition and \textbf{C}ircle-\textbf{S}upervision, we propose a novel \textbf{P}oisson \textbf{F}low \textbf{G}enerative \textbf{M}odel (JCCS-PFGM) to address the aforementioned challenge for progressive low-dose multiphase CECT reconstruction. Our approach takes advantage of features correlations among multi-phase images and utilizes the powerful capability of an adapted PFGM to progressively reduce the scanning radiation dose of CECT to the ultra low level and achieve a high image quality with uncompromising clinical utilities. This approach promises to enable low-radiation risk of the multiple CECT which is scanned in a short time, and yet deliver excellent multiphase CECT images. The main contributions of JCCS-PFGM can be summarized as:
1) an effectively progressive low-dose reconstruction scheme is developed to leverage the imaging consistency and radiocontrast evolution along former-latter phases, so that enormously reduces the radiation dose needs and improve the reconstruction effect, even for the latter-phase scanning with extremely low dose;
2) a newly-designed circle-supervision strategy is proposed in PFGM to enhance the refactoring capabilities of normalized poisson field learned from the latent perturbed space to the specified CT image space, so that boosts the explicit reconstruction for noise reduction;
3) a novel joint condition mechanism is designed to explore correlation between former and current phases, so that extracts the complementary information for current noisy CECT and guides the reverse process of diffusion jointly with multiphase condition for optimal recovery of clinically relevant structures.
 %  结合多期间关系及PFGM 的。。。，逐层多期CT 剂量降至极低，并重建  noise redcution and structie maintaince

\begin{figure}[t]
\centering
\includegraphics[width=1\textwidth,height=0.4\textwidth]{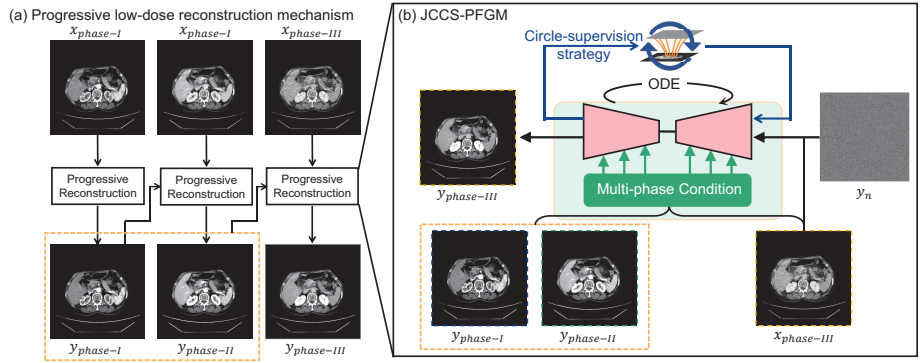}
\vspace*{-0.5\baselineskip}
\caption{JCCS-PFGM  is designed for progressive low-dose multiphase CECT reconstruction. It consists of the three key elements: progressive low-dose reconstruction scheme, circle-supervision strategy, and joint condition mechanism.
}\label{fig2}
\vspace*{-1.5\baselineskip}
\end{figure}

\section{Methodology}
\vspace*{-0.5\baselineskip}
As shown in Fig.~\ref{fig2}, the proposed JCCS-PFGM is progressively performed on multiphase low-dose CECT to reduce radiation risk in multiple CT imaging and target high-quality reconstruction for suppressed noise and structural fidelity. It is conducted with three special designs:
1) the progressive low-dose reconstruction scheme (detailed in Sect. 2.1) reasonably utilizes the consistency through the multiphase CECT imaging process phase-by-phase, reducing the image noise in the current phase based on the prior knowledge obtained from previous phase reconstructions;
2) the circle-supervision strategy (detailed in Sect. 2.2) embedded in PFGM makes further self-inspection on normal poisson field prediction, by penalizing any deviation from the current gradient flow; %by 使其感知 域的要求分布？.
and 3) the joint condition mechanism (detailed in Sect. 2.3)  integrates the multiphase consistency during the evolution of CT contrast material in the reverse diffusion process via fusing the complementary information from previous phases into the current phase of low-dose CECT.

\subsection{Progressive low-dose reconstruction scheme}
The progressive low-dose reconstruction scheme effectively promotes high-level base from former-phase to latter-phase for successively multiphase CECT reconstruction. Instead of distributing a total radiation dose evenly among all the phases, earlier phases use more radiation than latter phases because stronger prior information becomes available in the latter cases. Based on this idea, the inherent consistency traceable along different phases is encouraged for low-dose multi-phase CECT, which means a less multiphase reconstruction burden than that in the case of independent
reconstruction at each phase.

%effectively utilizes the inherent consistency along the multiphase CECT imaging, and promotes
%promotes the multiphase CECT

As show in Fig.~\ref{fig2}(a), our reasonable-designed progressive low-dose CECT reconstruction scheme assigns the dose from relatively high to low along the causal multiphase of phases I, II and III. With this scheme, the reconstruction at an earlier phase acquire more scanning information, benefit from relatively higher dose. And the latter phase is granted with much more reliable priori knowledge, benefit from the consistently traceable former-phase reconstruction. Denote the low-dose CECT images at phases I, II and III as $x_{phase-I}$, $x_{phase-II}$ and $x_{phase-III}$, our progressive reconstruction procedure is formulated as follows:

\begin{equation}\small
\left\{
\begin{split}
&y_{phase-I}=\mathcal{R}_1(x_{phase-I})\\
&y_{phase-II}=\mathcal{R}_2(x_{phase-II},y_{phase-I})\\
&y_{phase-III}=\mathcal{R}_3(x_{phase-III},[y_{phase-I},y_{phase-I}])\\
\end{split}\right.
\end{equation}
where $y_{phase-I}$, $y_{phase-II}$ and $y_{phase-III}$ represent  the phase-specific reconstruction results respectively, $\mathcal{R}_1(\cdot)$, $\mathcal{R}_2(\cdot)$ and $\mathcal{R}_3(\cdot)$ are the corresponding reconstruction models respectively.

\begin{figure}[t]
\centering
\includegraphics[width=0.8\textwidth,height=0.47\textwidth]{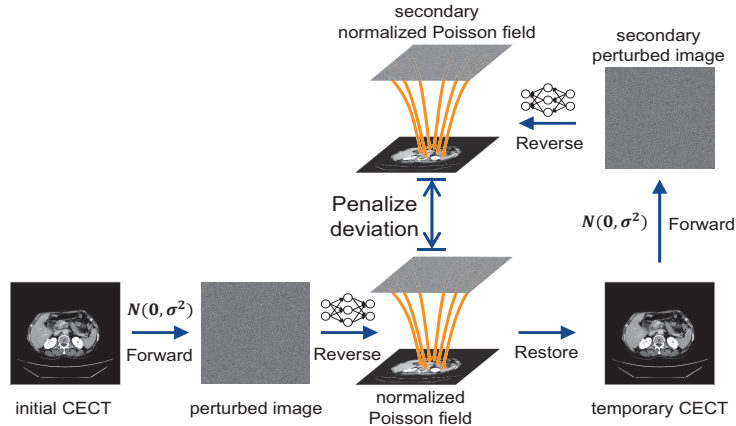}
\vspace*{-0.5\baselineskip}
\caption{The circle-supervision strategy robustly boosts the refactoring capabilities of normalized Poisson field learned by PFGM, by penalizing any deviation from the current gradient flow.
}\label{fig3}
\vspace*{-1.5\baselineskip}
\end{figure}

\subsection{Circle-supervision strategy strategy embedded in PFGM}

The circle-supervision strategy robustly boosts the refactoring capabilities of normalized Poisson field learned by PFGM. So that it further promotes the explicit reconstruction for noise reduction, instead of just doing a generic CT image generation.

PFGM is good at mapping a uniform distribution on a high-dimensional hemisphere into any data distribution \cite{{PFGM2022}}. It is inspired by electrostatics, and interpret initial image data points as electrical charges on the $z = 0$ hyperplane. Each initial image is able to be gradually transformed into a uniform distribution on the hemisphere when its radius $r\rightarrow \infty$. Then, it estimates the normalized Poisson field with a deep neural network (DNN), and alternatively uses a backward ordinary differential equation (ODE) solver for computationally efficient sampling.

The purpose of our circle-supervision strategy is to refine the normalized Poisson field which reflects the mapping direction from the latent perturbed space to the specified CECT image space. It promotes a precise perception on the target initial CECT imaging, facilitates gradient learning, and enhances crucial field components. As shown in Fig. ~\ref{fig3}, after randomly perturbing images in the forward process, the DNN estimates the normalized Poisson field $\phi_1$. Then, according to the normalized field calculation in the forward process, the Poisson field is returned with denormalization operation, and further temporarily restore the perturbed image into the initial CECT image space. The secondary diffusion process is conducted with same perturbtion in forward process and the same DNN in reverse process. Finally, the normal Poisson field is estimated of the secondary diffusion $\phi_2$. The deviation between $\phi_1$ and $\phi_2$ is penalized to boost the refactoring capabilities. Besides the temporary CECT is also yield in the secondary diffusion to enhance the robustness.

\begin{figure}[t]
\centering
\includegraphics[width=0.85\textwidth,height=0.49\textwidth]{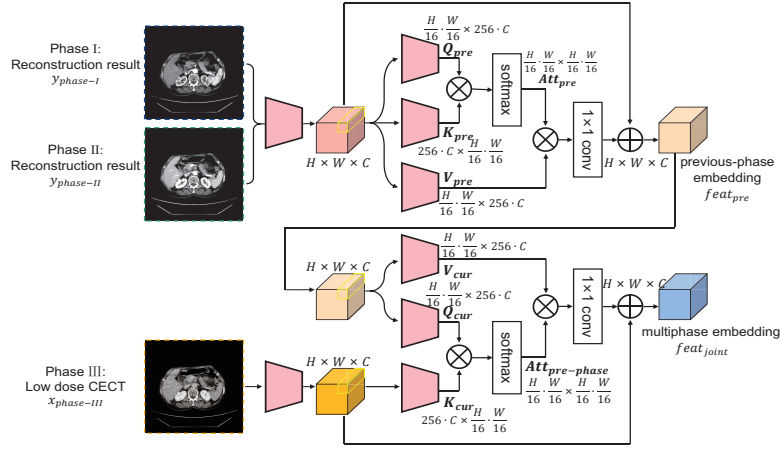}
\vspace*{-0.5\baselineskip}
\caption{The joint condition comprehensively fuses the consistency and evolution from
the previous phases to enhance the current ultra low-dose CECT image quality (taking the phase III CECT reconstruction for an example). It is composed of self-fusion among previous phases for consistency, and cross-fusion between the previous and current phases to capture the evolution process.
}\label{fig4}
\vspace*{-1.5\baselineskip}
\end{figure}

\subsection{Joint condition fusing multiphase consistency and evolution}

The joint condition comprehensively fuses the consistency and evolution from the previous phases to enhance the current ultra low-dose CECT. With this multiphase fusion, the reverse diffusion process can obtain increasingly more guidance to track the radiocontrast evolution dynamics for structure maintenance.

As shown in Fig. ~\ref{fig4}, the joint condition consists of two parts: the self-fusion among previous phases for consistency and the cross-fusion between previous and current phases to follow the evolution process.
1) For the self-fusion among previous phases, the reconstruction results from the previous phase I $y_{phase-II}$ and phase II $y_{phase-II}$ are combined into the feature domain. Then, the key map $K_{pre}$, query map $Q_{pre}$ and value map $V_{pre}$ are generated with further encoder and permutation. $K_{pre}$ and $Q_{pre}$ together establish the correlation weight, i.e., attention map $Att_{pre}$, among former phases combination to learn the inherent consistency and obtain the value map $V_{pre}$ to extract the consistent information which is finally added on the first feature representation for the previous-phases embedding $feat_{pre}$. The procedure is formulated as follows:
\begin{equation}\small
\left\{
\begin{split}
&K_{pre}=E_{K-pre}(E_{pre}([y_{phase-I},y_{phase-II}]))\\
&Q_{pre}=E_{Q-pre}(E_{pre}([y_{phase-I},y_{phase-II}]))\\
&V_{pre}=E_{V-pre}(E_{pre}([y_{phase-I},y_{phase-II}]))\\
&feat_{pre} = Conv(Softmax(Q_{pre}K_{pre}/\sqrt{d})V_{pre}) + E_{pre}([y_{phase-I},y_{phase-II}])
\end{split}\right.
\end{equation}
where $E_{pre}(\cdot)$,$E_{K-pre}(\cdot)$,$E_{Q-pre}(\cdot)$ and $E_{V-pre}(\cdot)$ are the corresponding encoders, $Softmax(\cdot)$ denotes Softmax function, and $Conv((\cdot)$ represents $1\times1$ convolution.

2) For the cross-fusion between the previous and current phases, it uses $feat_{pre}$ to generate query map $Q_{cur}$ and value map $V_{cur}$. The current phase III low-dose CECT $x_{phase-III}$ is encoded to produce key map $K_{cur}$. Thus the evolution from the previous to current phase is sensed between $K_{cur}$ and $Q_{cur}$ by attention map $Att_{cur}$. Then, the complimentary evolution from previous phases is extracted from the value map $V_{cur}$ with $Att_{cur}$, and added into the current phase embedding to form the multi-phase embedding. The procedure is formulated as:
\begin{equation}\small
\left\{
\begin{split}
&K_{cur}=E_{K-cur}(E_{cur}(x_{phase-III}))\\
&Q_{cur}=E_{Q-cur}(feat_{pre})\\
&V_{cur}=E_{V-cur}(feat_{pre})\\
&feat_{cur} = Conv(Softmax(Q_{cur}K_{cur}/\sqrt{d})V_{cur}) + E_{cur}(x_{phase-III})
\end{split}\right.
\end{equation}
where the new variables are similarly defined.

\begin{table}[t]
\scriptsize
\caption{The quantitative analysis of the proposed method under different configurations. (PLDRM: Progressive low-dose reconstruction mechanism with direct concatenation, CS: Circle-supervision strategy)}
\label{T1}
\begin{tabular}{c|ccc|ccc|ccc}
\hline
\multirow{2}{*}{Method} & \multicolumn{3}{c|}{Phase I (30\%dose)} & \multicolumn{3}{c|}{Phase II (15\%dose)} & \multicolumn{3}{c}{Phase III (5\%dose)} \\
                        & MAE($\downarrow$)     & PSNR ($\uparrow$)    & SSIM($\uparrow$)      & MAE($\downarrow$)     & PSNR ($\uparrow$)     & SSIM($\uparrow$)      & MAE($\downarrow$)     & PSNR($\uparrow$)      & SSIM($\uparrow$)      \\ \hline
PFGN                    &7.80HU     &47.4dB       & 97.3\%         & 10.3HU        &  44.8dB        &  97.1\%        & 16.0HU        & 42.5dB         & 96.8\%         \\
+PLDRM                  &7.80HU     &47.4dB       & 97.3\%         & 9.7HU        &  45.3dB       &  97.7\%        &  15.3HU       &  43.1dB        & 97.2\%         \\
+CS                     &7.23HU     &47.9dB       &98.0\%          & 9.2HU        &  45.6dB        & 98.0\%         &  14.8HU       &  43.8dB        & 97.5\%         \\
\textbf{Ours}           & \textbf{6.12HU}    & \textbf{48.4dB}    & \textbf{98.8\%}    &\textbf{8.49HU}         &\textbf{46.0dB}          &  \textbf{98.7\%}        &   \textbf{14.4HU}      &   \textbf{44.7dB}       &  \textbf{98.0\%}           \\ \hline
\end{tabular}
\end{table}

\section{Experiments}
\vspace*{-0.5\baselineskip}
\subsection{Materials and Configurations}
\vspace*{-0.5\baselineskip}
A clinical dataset consists 38496 CECT images from 247 patient were used in the experiment. Each patient went through a triple-phase CECT examination. We randomly divide the dataset into a training set of 123 patients, a validation set of 49 patients, and a testing set of 75 patients. The basic DNN used in reverse process is same as PFGM. The joint condition was introduced in DNN by making cross attention with DNN feature. The mean absolute error (MAE), SSIM and peak signal-to-noise-ratio(PSNR) were used as the performance metrics. The corresponding multiphase low-dose CECT was simulated by validated photon-counting model that incorporates the effect of the bowtie filter, automatic exposure control, and electronic noise \cite{Yu2012}%, as:
%\begin{equation}
%\small
%P_{B} = P_{A} + \sqrt{\frac{1-a}{a}\frac{exp(P_A)}{N_{0A}}(1+\frac{1+a}{a}\frac{N_eexp(P_A)}{N_{0A}})x}
%\end{equation}

\begin{table}[t]
\centering
\scriptsize
\caption{The quantitative analysis of the proposed method compared with the existing methods.}
\label{T2}
\begin{tabular}{c|ccc|ccc|ccc}
\hline
\multirow{2}{*}{Method} & \multicolumn{3}{c|}{Phase I (30\%dose)} & \multicolumn{3}{c|}{Phase II (15\%dose)} & \multicolumn{3}{c}{Phase III (5\%dose)} \\
                        & MAE($\downarrow$)     & PSNR($\uparrow$)    & SSIM($\uparrow$)     & MAE($\downarrow$)     & PSNR($\uparrow$)     &                         SSIM($\uparrow$)     & MAE($\downarrow$)     & PSNR($\uparrow$)     & SSIM($\uparrow$)     \\ \hline
FBP                &15.78HU         &40.1dB         & 93.9\%         &  21.5HU       &   37.6dB       &  89.2\%        &  34.9HU       &  33.7dB        &  77.9\%        \\
RED-CNN                   & 7.76HU        &  47.0dB       &  97.7\%        &  11.8HU       & 44.2dB         &96.2\%          & 18.6HU        &  42.2dB   &     95.4\%        \\
CLEAR                   &  8.61HU       &   45.9dB      &   96.2\%       &  10.1HU       &  45.4dB        & 97.4\%         & 19.7HU        &    41.5dB      &  95.9\%        \\
DDPNet                  &  8.07HU       &  46.8dB       &   97.5\%       &  10.4HU       & 45.8dB         & 98.0\%         &  16.8HU       & 42.7dB         & 97.1\%         \\
\textbf{Ours}                    & \textbf{6.12HU}    & \textbf{48.4dB}    & \textbf{98.8\%}    &\textbf{8.49HU}         &\textbf{46.0dB}          &  \textbf{98.7\%}        &   \textbf{14.4HU}      &   \textbf{44.7dB}       &  \textbf{98.0\%}        \\ \hline
\end{tabular}
\end{table}

\subsection{Results and Analysis}

\subsubsection{Overall performance.}As the last column shown in Tables~\ref{T1} and \ref{T2}, the proposed JCCS-PFGM gains high-quality multiphase low-dose CECT reconstruction with MAE down to 9.67 HU for information recovery, PSNR up to 46.3dB for noise reduction, and SSIM up to 98.5\% for structure recovery, as averagely on all testing images in phases I, II and III.

\subsubsection{Ablation study.} As shown in Table~\ref{T1}, the proposed JCCS-PFGM approach achieved error decrease of 1.23HU, PSNR increase of 1.05dB, and SSIM improvement by 1.06\%, on average, compared to the basic PFGM with only current-phase condition, and the various configurations by successively adding progressive low-dose reconstruction scheme and circle-supervision strategy. It justifies each and every key algorithmic element in our proposed approach. Especially for phase III with ultra low-dose of 5\%, it gets great improvement.

\begin{figure}[t]
\centering
\includegraphics[width=0.8\textwidth]{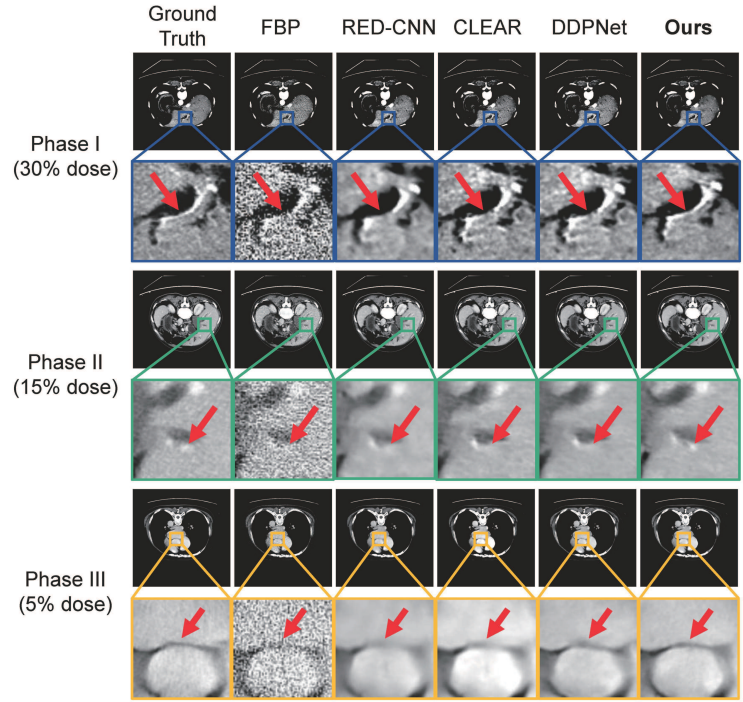}
\caption{Visual comparison with existing competing methods. It can be seen that the proposed JCCS-PFGM effectively keeps structural subtitles in all
phases, where the variable dose assignment scheme is as follows: phase I 30\% of the total nominal dose, phase II 15\% and phase III 5\%.
}\label{fig5}
\end{figure}

\subsubsection{Comparison with  competing methods.}
As shown in Table~\ref{T2}, the proposed JCCS-PFGM approach outperformed the competing methods including FBP, RED-CNN \cite{Chen2017_1}, CLEAR \cite{CLEAR2021} and DDPNet \cite{Ge2022}, with error decrease of 5.66 HU, PSNR increase of 3.62dB and SSIM improvement by 3.88\% on average. Visually,
Fig. \ref{fig5} illustrates that the result from JCCS-PFGM preserved tiny features in all the phases with the dose assignment scheme: phase I 30\% of
the total nominal dose, phase II 15\% and phase III 5\%.
In the enlarged ROI where the interpretation is difficult with original LDCT images, our method revealed key details, such as the vessel indicated by the red arrow, much better than the compared methods.

\section{Conclusion}
In this paper, we propose a cutting-edge JCCS-PFGM model for low-dose multiphase CECT. JCCS-PFGM creatively consists of the three key components:1) the progressive low-dose reconstruction scheme utilizes the consistency along the multiphase CECT imaging; 2)the circle-supervision strategy embedded in PFGM makes further self-inspection on normal poisson field prediction; 3) the joint condition integrates the multi-phase consistency and evolution in guiding the reverse process of diffusion. Extensive experiments have produced promising results quantitatively and qualitatively. fWe believe that with
further development our approach should have a great clinical potential in CECT applications.

\bibliographystyle{splncs04}
\bibliography{ref}

\end{document}